\documentclass[conference]{IEEEtran}

\usepackage{graphicx}
\usepackage{wrapfig}
\usepackage{amsmath}
\usepackage{url}
\usepackage{booktabs}
\usepackage{topcapt}
\usepackage{times}
\usepackage{textcomp}
\usepackage{sidecap}
\usepackage[]{mcode}
\usepackage{setspace}
\usepackage{multicol}
\usepackage{moresize}
\usepackage[normalem]{ulem}
\usepackage{newclude}
\usepackage{wrapfig}
\usepackage{graphicx} 
\usepackage{epsfig} 
\usepackage{mathptmx} 
\usepackage{times} 
\usepackage{amsmath} 
\usepackage{amssymb}  
\usepackage[]{algorithm2e}

\begin{document}

\title{Big Data Dimensional Analysis}

\author{\IEEEauthorblockN{Vijay Gadepally \& Jeremy Kepner}
\IEEEauthorblockA{MIT Lincoln Laboratory, Lexington, MA 02420}
\IEEEauthorblockA{\{vijayg, jeremy\}@ ll.mit.edu}}

\maketitle

\let\thefootnote\relax\footnote{This work is sponsored by the
  Assistant Secretary of Defense for Research and Engineering under
  Air Force Contract \#FA8721-05-C-0002.  Opinions, interpretations,
  recommendations and conclusions are those of the authors and are not
  necessarily endorsed by the United States Government \\
  978-1-4799-6233-4/14/\$31.00 \textcopyright 2014 IEEE}

\begin{abstract}
The ability to collect and analyze large amounts of data is a growing
problem within the scientific community. The growing gap between data
and users calls for innovative tools that address the challenges faced
by big data volume, velocity and variety.  One of the main challenges
associated with big data variety is automatically understanding  the 
underlying structures and patterns of the data. Such an understanding is required as a
pre-requisite to the application of advanced analytics to the
data. Further, big data sets often contain anomalies and errors that
are difficult to know {\it a priori}. Current approaches to understanding
data structure are drawn from the traditional database ontology design.
These approaches are effective, but often require too much human involvement to be
effective for the volume, velocity and variety of data encountered by big data systems.
Dimensional Data Analysis (DDA) is a proposed technique that allows big data analysts to quickly understand the
overall structure of a big dataset, determine anomalies. DDA
exploits structures that exist in a wide class of data to quickly determine
the nature of the data and its statical anomalies.  DDA leverages existing
schemas that are employed in big data databases today.  This paper presents
DDA, applies it to a number of data sets, and measures its performance.  The overhead
of DDA is low and can be applied to existing big data systems without
greatly impacting their computing requirements.
\end{abstract}

\begin{IEEEkeywords}
Big Data, Data Analytics, Dimensional Analysis
\end{IEEEkeywords}

\section{Introduction}

The challenges associated with big data are commonly referred to as
the 3 V's of Big Data - Volume, Velocity and
Variety~\cite{laney20013d}. The 3 V's provide a guide to the largest outstanding challenges associated with
working with big data systems. Big data \textbf{volume} stresses the storage,
memory and compute capacity of a computing system and requires access
to a computing cloud. The \textbf{velocity} of big data velocity stresses the rate at which data can
be absorbed and meaningful answers produced. Big data \textbf{variety}
makes it difficult to develop algorithms and tools that can address
that large variety of input data.

The MIT Supercloud infrastructure~\cite{reuther2013llsupercloud} is designed to address the challenge of big data volume.
To address big data velocity concerns, MIT Lincoln Laboratory worked
with various U.S. government agencies to develop the Common Big Data
Architecture and its associated Apache Accumulo database. Finally, to address big data variety problems, MIT
Lincoln Laboratory developed the D4M technology and its associated
schema~\cite{kepner2012dynamic} that is widely used across Accumulo community. 

While these techniques and technologies continue to evolve with the 
increase in each of the V's of big data, analysts who work with data
to extract meaningful knowledge have realized that the ability to quantify low level
parameters of big data can be an important first step in
an analysis pipeline. For example, in the case of machine learning,
removing extraneous dimensions or erroneous records allows the
algorithms to focus on meaningful data. Thus, the first step of a machine learning analyst is manually cleaning the big dataset or performing dimensionality
reduction through techniques such as random
projections~\cite{bingham2001random} or
sketching~\cite{indyk1998approximate}. Such tasks require a coherent
understanding of the data set
which can also provide insight into any weaknesses that may be
present in a data set. Further, detailed analysis of each data set
is required to determine any internal patterns that may exist.

The process for analyzing a big data set can often be summarized as follows:
\begin{enumerate}
\item Learn about data structure through Dimensional Data Analysis (DDA);
\item Determine background model of big data;
\item Use data structure and background model for feature extraction,
  dimensionality reduction, or noise removal;
\item Perform advanced analytics; and
\item Explore results and data.
\end{enumerate}

These steps can be adapted to a wide variety of data and borrow
heavily from the processes and tools developed for the signal
processing community. The first two steps in this process provide a
high level view of a given data set - a very important step to ensure
that data inconsistencies are known prior to complex analytics that
may obscure the existence of noise.  Traditionally,  this view
is obtained as a byproduct of standard database ontology techniques, whereby
a database analyst or data architecture examines the data in detail prior to assembling
the database schema.  In big data systems,  data can change quickly or whole new classes of data
can appear unexpectedly and it is not feasible for this level of analysis to be employed.
The D4M (d4m.mit.edu) schema addresses part of this problem by allowing a big data system to
absorb and index a wide range of data with only a handful of tables that are consistent
across different classes of data.  An added byproduct of D4M schema is that common
structures emerge that can be exploited to quickly or automatically characterize data.

Dimensional Data Analysis (DDA) is a technique to learn about data structure
and can be used as a first step with a new or unknown big data
set. This technique can be used to gain an understanding of
corpus structure, important dimensions, and data corruptions (if
present).

The article is organized as follows. Section~\ref{technology} describes the MIT SuperCloud
technologies designed to mitigate the challenges associated with the 3
V's of big data. Section~\ref{dimanalysis} provides the mathematical and
application aspects of dimensional analysis. In order to illustrate
the value of dimensional analysis, two applications are described in
Section~\ref{examples} along with performance measurements for the applications of this approach. Finally, section~\ref{conclusions} concludes the article and discusses future work.

\section{Big Data and MIT SuperCloud}
\label{technology}

The growing gap between data generation and
users has influenced the
movement towards cloud computing that can offer centralized large scale
computing, storage and communication networks. Currently, there are four multibillion dollar ecosystems that dominate the cloud computing
environment: enterprise clouds, big data clouds, SQL database clouds,
and supercomputing clouds. The MIT Supercloud
infrastructure was developed to allow the co-existence of all four
cloud ecosystems on the same hardware without sacrificing performance
or functionality. The MIT Supercloud uses the Common Big Data
Architecture (CBDA), an architectural abstraction that describes the flow of information within such
systems as well as the variety of users, data sources, and system
requirements. The CBDA has gained widespread adoption and uses the NSA
developed Accumulo database~\cite{apacheaccumulo} that
has demonstrated high performance capabilities (capable of hundreds millions of
database entries/second) and has been used in a variety of
applications~\cite{byun2012driving}. These technologies and other
tools discussed in this section are used to develop the dimensional
analysis technique.

\subsection{Big Data Pipeline}
\begin{figure*}
\vspace{-5pt}
\centerline{
\includegraphics[width=6in]{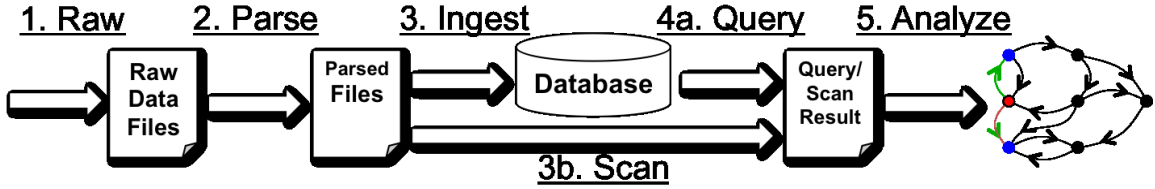}
}
\caption{Working with big data usually implies working with these 5 steps.}
\vspace{-10pt}
\label{pipeline}
\end{figure*}

A big data pipeline is a distilled view
of the CBDA meant to describe the system components and interconnections involved in most big
data systems. The pipeline in Figure~\ref{pipeline} has
been applied to diverse big data problems such as health care, social media, defense
applications, intelligence reports, building management systems, etc.

The generalized five step system was created after observing numerous
big data systems in which the following steps are performed
(application specific names may differ, but the component's purpose is
usually the same):

\begin{enumerate}
\item Raw Data Acquisition: Retrieve raw data from external sensors or
  sources.
\item Parse Data: Raw data is often in a format which needs to be
  parsed. Store results on distributed filesystem.
\item Ingest Data: If using a database, ingest parsed data into
  database.
\item Query or Scan for Data: Use either database or filesystem to
  find information.
\item Analyze Data: Perform complex analytics, visualizations,
  etc. for knowledge discovery.
\end{enumerate}

\subsection{Associative Arrays}

In order to perform complex analytics on databases, it is necessary to
develop a mathematical formulation for big data types. Associations between multidimensional entities (tuples) using
number/string keys and number/string values can be stored in data
structures called associative arrays. Associative arrays are used as
the building block for big data types and consists of a collection of
key value pairs. 

Formally, an associative array \textbf{A} with possible keys
$\{k_{1}, k_{2}, ..., k_{d}\}$, denoted as \textbf{A(k)} is a partial
function that maps two spaces \textbf{$S^{d}$} and \textbf{$S$}:

\begin{eqnarray}
\nonumber A(k) : S^{d}\rightarrow S \mbox{         } k = (k^{1}, ... ,
k^{d}) \\
\nonumber k^{i} \in S^{i}
\end{eqnarray}

Where $A(k)$ is a partial function from $d$ keys to one value where:

\begin{eqnarray}
\nonumber A(k_{i}) = \left\{\begin{matrix}
v_{i} & \\ 
\phi  & otherwise 
\end{matrix}\right.
\end{eqnarray}

Associative arrays support a variety of linear algebraic operations
such as summation, union, intersection, multiplication. Summation of two associative arrays, for example, that do not have any common row or column key performs a concatenation.  In the D4M schema a table in the Accumulo database is an associative array.

\subsection{D4M \& D4M Schema}
\label{d4m}

NoSQL databases such as Accumulo have become a popular alternative to
traditional Database Management Systems such as SQL. Such databases
require a database schema which can be difficult due to big data
variety. Big data variety challenges the tools and algorithms developed to
process big data sets. The promise of big data is the ability to correlate diverse
and heterogeneous data sources to reduce the time to insight. 
Correlating this data requires putting each
format into a common frame of reference so that like entities can be
compared. D4M~\cite{kepner2012dynamic} allows vast quantities of
highly diverse data to automatically be ingested into a simple common schema
that allows every unique element to be quickly queried and
correlated. 

Within the CBDA, the D4M environment is used to support prototyping of analytic solutions for big data
applications. D4M applies the concepts of linear algebra and signal
processing to databases through associative arrays; provides a data
schema capable of representing most data; and provides a low barrier
to entry through a computing API implemented in MATLAB and GNU Octave.

The D4M 2.0 Schema~\cite{kepner2013d4m}, provides a four table solution
that can be used to represent most data values. The four table solution allows diverse data to be stored in a
simple, common format with just a handful of tables that can
automatically be generated from the data with no human
intervention. From the schema described in~\cite{kepner2013d4m}, a dense database can be
converted by ``exploding'' each data entry into an associative array
where each unique column-value pair
is a column. Once in sparse matrix form, the full machinery of linear
algebraic graph processing \cite{kepner2011graph, kepner2013taming} and detection theory can be
applied. For example, multiplying two associative arrays correlates the two.

\section{Dimensional Data Analysis}
\label{dimanalysis}

Dimensional Data Analysis (DDA) provides a principled way to develop a coherent
understanding of underlying data structures, data inconsistencies,
data patterns, data formatting, etc. Over time, a database may develop
inconsistencies or errors, often due to a variety of reasons. Further,
it is very common to perform advanced analytics on a database,
often looking for small artifacts in the data which may be of
interest. In these cases, it is important for a user to understand the
information content of their database.

\begin{figure*}
\centerline{
\includegraphics[width=5.5in]{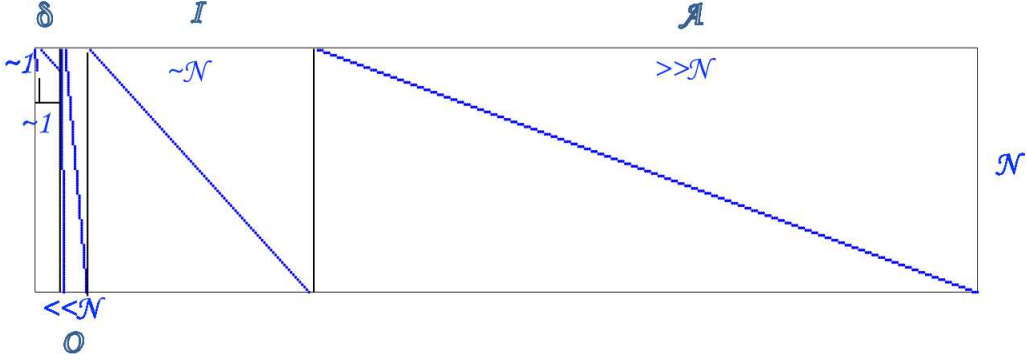}
}
\vspace{-8pt}
\caption{A database can be represented as the sum (concatenation) of a
series of sub associative arrays that correspond to different ideal or
vestigial arrays}
\vspace{-8pt}
\label{fi:idealsumarray}
\end{figure*}

First, it is necessary to define the components of a database in
formal terms.

A database can be represented by a model that is described as a
sum of sparse associative arrays, Figure~\ref{fi:idealsumarray}. Consider a database
\textbf{\underline{E}} represented as the sum of sparse sub-associative
arrays \textbf{$E_{i}$}:

\vspace{-5pt}
\begin{eqnarray}
\nonumber \textbf{E} = \sum_{1}^{n} E_{i}
\end{eqnarray}

Where $i$ corresponds to different entities that comprise the $n$
entities of \textbf{\underline{E}}. Each \textbf{$E_{i}$} has the
following properties/definitions:

\begin{itemize}
\item $N$ is the number of rows in the whole database (number of rows
  in associative array \textbf{\underline{E}}).
\item $N_{i}$ is the number of rows in database entity $i$ with
  at least one value (number of rows in associative array \textbf{$E_{i}$}).
\item $M_{i}$ is the number of unique values in database column $i$
  (number of columns in associative array \textbf{$E_{i}$}).
\item $V_{i}$ is the number of values in database column $i$ (number
  of non zero values in associative array \textbf{$E_{i}$}).
\end{itemize}

With these definitions, the following global sums hold:

\begin{eqnarray}
\nonumber N \leq \sum_{i} N_{i}, \mbox {      ,} \forall i \\
\nonumber M = \sum_{i} M_{i} \mbox {     , }  \forall i \\
\nonumber V = \sum_{i} V_{i} \mbox {     , }  \forall i 
\end{eqnarray}

where $N$, $M$, and $V$ correspond to the number of rows, columns and
values in database \textbf{\underline{E}} respectively.

Theoretically, each sub-associative array (\textbf{$E_{i}$}) can be typed as
\textit{ideal} or \textit{vestigial} arrays depending on the
properties of this sub-associative array:

\begin{itemize}
\item Identity (\textbf{I}): Sub-associative array \textbf{$E_{i}$} in
  which the number of rows and columns are of the same order:
\begin{eqnarray} 
\nonumber N_{i} \sim  M_{i} 
\end{eqnarray}

\item Authoritative (\textbf{A}): Sub-associative array \textbf{$E_{i}$} in
  which the number of rows is significantly smaller than the number of
  columns:
 \begin{eqnarray} 
\nonumber N_{i} \ll  M_{i} 
\end{eqnarray}

\item Organizational (\textbf{O}): Sub-associative array \textbf{$E_{i}$} in
  which the number of rows is significantly greater than the number of
  columns:
 \begin{eqnarray} 
\nonumber N_{i} \gg  M_{i} 
\end{eqnarray}

\item Vestigial (\textbf{$\delta$}): Sub-associative array
    \textbf{$E_{i}$} in which the number of rows and columns are
    significantly small
 \begin{eqnarray} 
\nonumber N_{i} \sim 1 \\
\nonumber M_{i} \sim 1
\end{eqnarray}
\end{itemize}

Conceptually, data collection for each of the entities is intended to
follow the structure of ideal models. However, due to
inconsistencies and changes over time, they may develop vestigial
qualities or differ from the intended ideal array. By comparing a given sub associative array to the structures
described above, it is possible to learn about a
given database and recognize inconsistencies or errors.

\subsection{Performing DDA}

Consider a database \textbf{\underline{E}}. In a real system,
\textbf{\underline{E}} is a large sparse associative array representation of all
the data in a database using the schema described in the previous
sections. Suppose that
\textbf{\underline{E}} is made up of $k$ entities, such that:

\begin{eqnarray}
\nonumber \textbf{E} = \sum_{1}^{k} E_{i} 
\end{eqnarray}

In a real database, these entities typically relate to various
dimensions in the dataset. For example, entities may be time stamp,
username, building id number, etc. Each of the associative arrays
corresponding to $E_{i}$ is referred to as a sub-associative
array. Dimensional analysis compares the structure of each $E_{i}$
with the intended structural model. This process
consists of the steps described in Algorithm~\ref{algo1}. 

\begin{algorithm}
 \KwData{DB represented by sparse associative array E}
 \KwResult{Dimensions of sub-associative arrays corresponding to entities}
 \For{entity i in k}{
  read sub-associative array $E_{i}\in E$\;
  \eIf{number of rows in $E_{i} \ge 1$}{
   number of rows in $E_{i}$ = $N_{i}$\;
   number of unique columns in $E_{i}$=$M_{i}$\;
   number of values in $E_{i}$=$V_{i}$\;
   }{
   go to next entity\;
  }
 }
\caption{Dimensional Analysis Algorithm}
\label{algo1}
\end{algorithm}

Using the algorithm above, let the dimensions of each sub associative array ($E_{i}$) be
contained in the 3-tuple ($N_{i}$, $M_{i}$, $V_{i}$) corresponding to
the number of rows, columns and values in each sub associative array
which corresponds to a single entity.

\subsection{Using DDA Results}

Once the tuples corresponding to each entity is collected for a
database \textbf{E}, one can compare the dimensions with the \textit{ideal} and \textit{vestigial}
arrays described in Section~\ref{dimanalysis} to determine the approximate intended structural model
for each entity.

Once the intended structural model for an entity is determined, it is possible to highlight
interesting patterns, anomalies, formatting, and
inconsistencies. For example:

\begin{itemize}
\item Authoritative (\textbf{A}): Important entity values (such as
  usernames, words, etc.) are highlighted by:
\begin{eqnarray} 
\nonumber E_{i} * 1_{Nx1} >1 \\
\nonumber 1_{1xN} * E_{i} >1
\end{eqnarray}

\item Identity (\textbf{I}): Misconfigured or non-standard entity values are highlighted by:
\begin{eqnarray} 
\nonumber E_{i} * 1_{Nx1} >1 \\
\nonumber 1_{1xN} * E_{i} >1 \\
\nonumber I_{NxN} - E_{i} \neq 0_{NxN}
\end{eqnarray}

\item Organizational (\textbf{O}): The mapping structure of a sub-associative array is highlighted by counts and correlations in which:
 \begin{eqnarray} 
\nonumber E_{i} * 1_{Nx1} \gg1 \\
\nonumber E_{i}^{T} * E_{j} >> 1 \\
\nonumber 1_{1xN} * E_{i} = 1
\end{eqnarray}

\item Vestigial (\textbf{$\delta$}): Erroneous or misconfigured
  entries can typically be determined by inspecting $E_{i}$.

\end{itemize}

The difference between a sub-associative array and an intended model
such as those above provide valuable information about failed
processes, corrupted or junk
data, non-working sensors, etc. Further, the actual dimensions of each
sub-associative array can provide information about the structure of a
database that enables a high level understanding of a particular data dimension.

\section{Application Examples}
\label{examples}

In this section, we provide two example data sets and the results
obtained through DDA. This section is meant to
illustrate the concepts described before.

\subsection{Geo Tweets Corpus }

Social media analysis is a growing area of interest in the big data
community. Very often, large amounts of data is collected through a
variety of data generation processes and it is necessary to learn
about the low level structural information behind such data. Twitter
is a microblog that allow up to 140 character ``tweets'' by
a registered user. Each tweet is published by Twitter and is available via a publicly
acessible API. Many tweets contain geo-tagged information if enabled
by the user. A prototype twitter
dataset containing 2.02 million tweets was used for the dimensional
analysis.

\subsubsection{Dimensional Analysis Procedure}

The process outlined in the previous section was used to perform
dimensional analysis on a set of Twitter data with the intent of
finding any anomalies, special accounts, etc. The database consists of 2.02 million rows and values
distributed across 10 different entities such as latitude, longitude,
userID, username, etc.

\begin{figure*}
\centerline{
\includegraphics[width=6in]{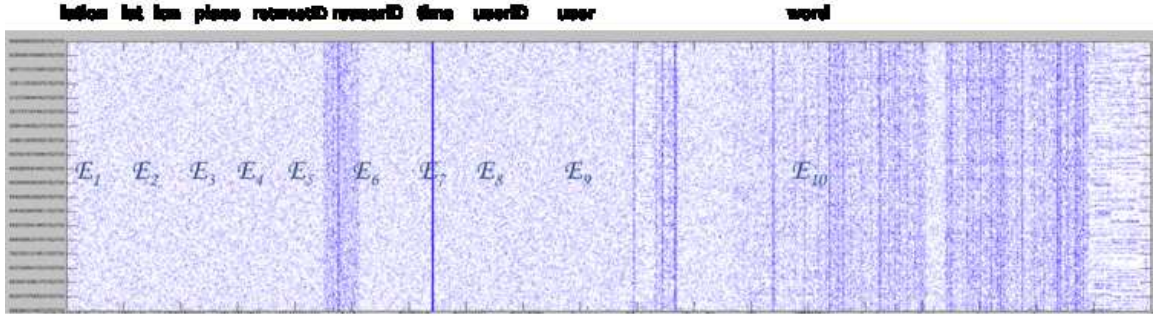}
}
\vspace{-5pt}
\caption{Associative Array representation of Twitter data. $E_{1},
  E_{2},...,E_{10}$ represent the concatenated associative arrays
  $E_{i}$ that constitute all the entities in the full dataset. Each
  blue dot corresponds to a value of 1 in the associative array representation.}
\label{twitterspy}
\end{figure*}

The associative array representation of the Twitter corpus is shown in
Figure~\ref{twitterspy}. The 10 \textit{dimensions} or entities of the database that make up
the full dataset are also shown.

\subsubsection{DDA Results}

Dimensional analysis of the dataset can be performed by performing
Algorithm 1 on each of the entities ($i$) in $k$ possible
entities. For example, $E_{7}$ = $E_{Time}$ which is the associative
array in which all of the column keys correspond to time stamps. Thus,
the triple ($N_{Time}, V_{Time}, M_{Time}$) is the number of entries with
a time stamp, number of time stamp entries in the corpus, and number of
unique time stamp values respectively. Performing Algorithm 1 on each
of the 10 entities yields the results described in Table~\ref{tweettable}.

\begin{table*}[ht]
\centering
\begin{tabular}{|c|c|c|c|c|}
\hline
Entity& $N_{i}$ & $V_{i}$ & $M_{i}$ &  Structure Type\\ \hline
latlon & 1624984 & 1625197 & 1506465 & Identity \\ \hline
lat & 1624984 & 1625192 & 1504469 & Identity \\ \hline
lon & 1625061 & 1625725 & 1504619 & Identity\\ \hline
place & 1741337 & 1741516 & 1504619 & Identity\\ \hline
retweetID & 636455 & 636644 & 627163 & Identity\\ \hline
reuserID & 720624 & 722148 & 676616 & Identity\\ \hline
time & 2020000 & 2020000 & 35176 & Organization\\ \hline
userID & 2020000 & 2020000 & 1711141 & Identity \\ \hline
user & 2020000 & 2020000 & 1711143 &  Identity\\ \hline
word & 1976746 & 17180314 & 7838862 & Authority\\ \hline
\end{tabular}
\vspace{5pt}
\caption{Dimensional Analysis performed on 2.02 million Tweets}
\label{tweettable}
\vspace{-8pt}
\end{table*}

Using the definitions defined in Section~\ref{dimanalysis}, we can quickly determine
important characters. For example, to find the most popular users, we can look at the
difference where $E_{user} * 1_{Nx1} >1$. Using D4M, this computation
can easily be performed with associative arrays to yield the most
popular users. Performing this analysis on the full 2.02 million tweet
dataset represented by an associative array \textbf{E}:

\begin{lstlisting} 
% Extract Associative Array Euser
 >>Euser = E(:,StartsWith('user|,')); 
%Add up count of all users
 >>Acommon = sum(Euser, 1);
%Display most common users
 >>display(Acommon>150);
   (1,user|SFBayRoadAlerts)     258
   (1,user|akhbarhurra)     177
   (1,user|attir_midzi)     159
   (1,user|verkehr_bw)     300
\end{lstlisting}

The results above indicate that there are 4 users who have greater
than 150 tweets in the dataset.

\subsubsection{DDA Performance}

\begin{figure}
\vspace{-5pt}
\centerline{
\includegraphics[width=4in]{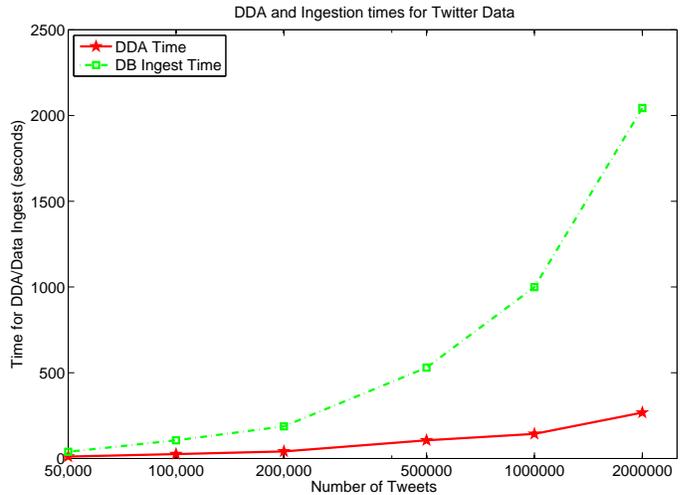}
}
\vspace{-5pt}
\caption{Relative performance between DDA and ingesting data.}
\vspace{-5pt}
\label{tweetperf}
\end{figure}

Ingesting data into a database is often an expensive and time consuming
process. One of the features of DDA is in the ability to potentially
reduce the amount of information that needs to be
stored. Figure~\ref{tweetperf} describes the relative time taken by
DDA compared to data ingest.

From this comparison, it is clear that DDA takes a fraction of the
time compared to ingest. By using DDA, one may be able to remove
entries that need to be ingested, thus reducing the overall ingest time.

\subsection{HPC Scheduler Log Files}

Another application in which dimensional analysis was tested is with
HPC scheduler log files. LLSuperCloud uses the Grid Engine scheduler for dispatching
jobs. For each job that is finished, an accounting record is written
to an accounting file. These records can be used in the future to
generate statistics about accounts, system usage, etc.  Each line in
the accounting file represents an individual job that has completed.

\subsubsection{DDA Procedure}

The process outlined in the previous section were used to perform
dimensional analysis on a set of SGE accounting data with the intent of
finding any anomalies, special accounts, etc. The database consists of
approximately 11.5 million entries with 27 entities each. A
detailed description of the entities in the SGE accounting file can be
found at: \cite{sgeaccountin}.

\begin{figure*}[t!]
\centerline{
\includegraphics[width=6in]{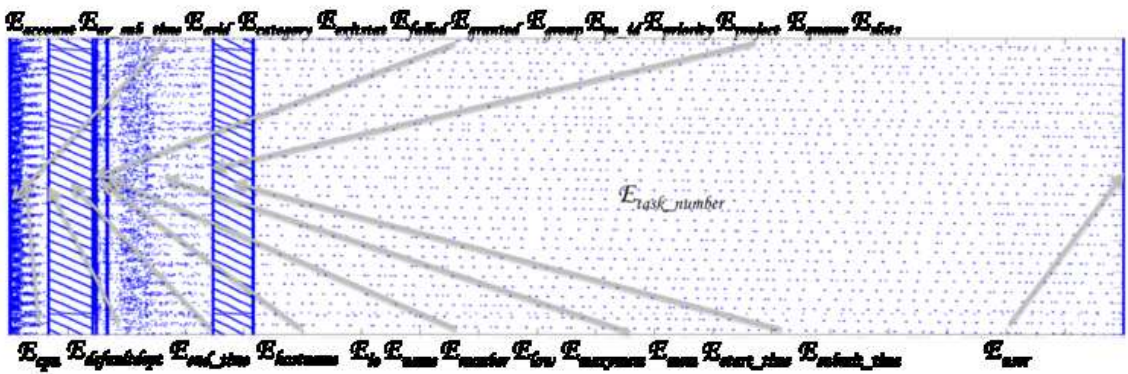}
}
\caption{Associative Array representation of SGE accounting data represent the concatenated associative arrays
  $E_{i}$ that make up the full dataset. }
\label{sgespy}
\end{figure*}

The associative array representation of the SGE corpus is shown in
Figure~\ref{sgespy}. The 27 ``dimensions'' of data that make up
the full data set are shown in figure~\ref{sgespy}.

\subsubsection{DDA Results}

After performing dimensional analysis on the SGE corpus, the results
are tallied for inspection. A subset of the results is shown in
Table~\ref{sgeresults}. It is interesting to note that there are many
accounting file entries that are not collected and have only default
values. For example, the field ``defaultdepartment'' contains only one
unique value in the entire dataset - ``default''. For an individual
wishing to perform more advanced analytics on the dataset, this is an
important result and can be used to reduce the dimensionality of each
data point.

\begin{table*}[ht]
\centering

\begin{tabular}{|c|c|c|c|c|}
\hline
Entity& $N_{i}$ & $V_{i}$ & $M_{i}$ &  Structure Type\\ \hline
Account & 11446187 & 11446187 &   1  & Vestigial \\ \hline
CPU Hours & 11446187 & 11446187  & 2752964 & Identity \\ \hline
Default Department & 11446187 & 11446187 &   1  & Vestigial\\ \hline
Job Name& 11446187 & 11446187  & 90491 & Organization \\ \hline
Job Number & 11446187 & 11446187 & 485212 & Identity \\ \hline
Memory Usage & 11446187 & 11446187 & 5241559 & Identity \\ \hline
Priority & 11446187 & 11446187 & 1 & Vestigial \\ \hline
Task Number & 11446187 & 11446187 & 7491889 & Identity \\ \hline
User Name & 11446187 & 11446187 & 8388 & Organization \\ \hline
\end{tabular}
\vspace{10pt}
\caption{Dimensional Analysis performed on ~11.5 million
  Sun Grid Engine accounting entries. Only selected entries are shown
  of the 27 total entries collected}
\label{sgeresults}
\end{table*}

A D4M code snipped to find the most common job names is shown below.

\begin{lstlisting} 
% Extract Associative Array Euser
 >>Ejobname = E(:,StartsWith('job_name|,'));
%Add up count of all users
 >>Acommon = sum(Ejobname, 1);
%Display most common users
 >>display(Acommon>1000000);
  (1,job_name|rolling_pipeline.sh)     2762791
  (1,job_name|run_blast.sh)     1256422
  (1,job_name|run_blast_parser.sh)     1162522
\end{lstlisting}

Interestingly, of the 27 dimensions of data in the SGE log files, 8 of
the entities are not actually recorded. This information can be very
important to one interested in performing advanced analytics on a
dataset in which nearly one third of the data is unchanging.

\section{Conclusions and Future Work}
\label{conclusions}

In this paper, we proposed a process to understand the structural
characteristics of a database called dimensional data
analysis. Using DDA, a researcher can learn
a great deal about the hidden patterns, structural characteristics and
possible errors of a large unknown database. DDA
consists of representing a dataset using associative arrays and
performing a comparison between the constituent associative arrays and
intended ideal database arrays. Deviations from the intended model can highlight
important details or incorrect information.

We recommend that the DDA technique be the first step of an
analytic pipeline. The common next steps in an analytic pipeline such as background modeling, feature
extraction, machine learning and visual analytics depend heavily on
the quality of input data.

Next steps to this work include developing an automated mechanism to
perform background modeling of big datasets, and application of
detection theory to big data sets.

\section*{Acknowledgment}

The authors would like to thank the LLGrid team at MIT Lincoln
Laboratory for their support and expertise in setting up the computing
environment.

\ifCLASSOPTIONcaptionsoff
  \newpage
\fi
\begin{spacing}{0.78}
\footnotesize
\bibliography{ieeehpec}
\end{spacing}
\vfill


\end{document}